\documentclass{optica-article}

\journal{oe} 

\usepackage[detect-weight=true, detect-family=true]{siunitx}
\usepackage{physics}
\usepackage{bm}

\begin{document}

\title{The effect of nonlinear lensing on the coupling of ultrafast laser pulses to hollow-core waveguides}

\author{Christian Brahms\authormark{1, *}}

\address{\authormark{1}School of Engineering and Physical Sciences, Heriot-Watt University, Edinburgh, EH14 4AS, UK}

\email{\authormark{*}Corresponding author: c.brahms@hw.ac.uk}

\begin{abstract}
Gas-filled hollow-core fibres are a flexible platform for the manipulation of ultrafast laser pulses through a variety of nonlinear optical effects. Efficient high-fidelity coupling of the initial pulses is very important for system performance. Here we study the effect of self-focusing in gas-cell windows on the coupling of ultrafast laser pulses into hollow-core fibres using (2+1)-dimensional numerical simulations. As expected, we find that the coupling efficiency is degraded and the duration of the coupled pulses changed when the entrance window is too close to the fibre entrance. The interplay of nonlinear spatio-temporal reshaping and the linear dispersion of the window create different results depending on the window material, pulse duration, and pulse wavelength, with longer-wavelength beams more tolerant of high intensity in the window. While shifting the nominal focus to compensate can restore some of the lost coupling efficiency, it improves the pulse duration only marginally. From our simulations we derive a simple expression for the minimum distance between the window and the HCF entrance facet. Our results have implications for the often space-constrained design of hollow-core-fibre systems, especially where the input energy is not constant. 
\end{abstract}

\section{Introduction}
Gas-filled hollow-core fibres (HCFs) are one of the most commonly used platforms for the spectral and temporal reshaping of ultrafast laser pulses. They offer several advantages, most importantly the confinement of light over extended interaction lengths; high damage threshold; pressure-tuneable nonlinearity and dispersion; and spatial filtering properties. HCFs of different types and widely varying dimensions have been used with a great variety of input laser pulses, ranging in energy from nanojoules~\cite{xiong_low-energy-threshold_2021} to the multi-millijoule regime~\cite{suda_generation_2005,nagy_generation_2019,fan_70_2021,nagy_generation_2020} and in pulse duration from nanoseconds~\cite{ippen_low-power_1970, couny_generation_2007} down to only a few femtoseconds\cite{schenkel_generation_2003-1, brahms_high-energy_2019}. 

With the exception of air-filled fibres\cite{mousavi_nonlinear_2018,debord_strong_2019,luan_efficient_2021} or in-vacuum beam transport~\cite{nagy_generation_2020}, maintaining the gas fill while allowing optical access requires windows on the gas cell containing the HCF. The nonlinear response of gases is much weaker than that of solid materials, and hence the intensity inside the fibre can be much higher than can be sustained by solid materials. This means that the windows cannot be located at the entrance and exit faces of the waveguide, but need to be far enough away for the beam entering or emerging from the waveguide to diverge sufficiently and the intensity to drop. The required distance between fibre and window is determined by two factors: firstly, a laser beam with very high intensity or fluence can damage the window material or any anti-reflection coatings. Laser damage can be avoided entirely by reducing the intensity or fluence sufficiently, and the required distance is easily calculated, assuming the damage threshold of the window is known. Secondly, the Kerr effect turns the window itself into a nonlinear lens, distorting the wavefront of the beam entering or exiting the waveguide. This is especially detrimental for the beam entering the fibre, as a change in the focusing condition can prevent the laser pulse from being coupled to the waveguide efficiently. This effect is always present to some degree, and it is less straightforward to determine how far away a window needs to be. A simple model based on a quadratic approximation to a Gaussian beam profile, which allows for the calculation of an effective focal length for the Kerr lens, has been tested experimentally for the coupling of high-intensity pulses into capillary targets for high-harmonic generation~\cite{ran_coupling_2020}. However, this approach allows for neither a detailed analysis of the spatio-temporal reshaping of the pulse by the nonlinear lens, nor the formulation of simple design guidelines to minimise or avoid these effects.

Here we study the effect of Kerr lensing in the entrance window of hollow-core fibre systems on the coupling of ultrafast laser pulses to the waveguide for a wide variety of conditions. Using radially symmetric (2+1)-dimensional numerical simulations, we calculate the field of the pulse coupled to the waveguide for a wide range of focusing conditions, HCF core sizes and initial pulse parameters. Unsurprisingly, we find that strong Kerr lensing can dramatically reduce the efficiency of coupling to the fundamental mode of the HCF. In addition, the variation of the optical power, and hence the strength of the Kerr lens, over the duration of the pulse leads to an effective pulse stretching. This combines with spectral compression or broadening, depending on the group-velocity dispersion of the window material. Pulses at longer wavelengths are less strongly affected by the Kerr lens and hence higher intensity can be tolerated. For a fixed peak power, the coupling efficiency can be largely restored by shifting the nominal focus to compensate; however, changing the pulse energy or peak power requires renewed adjustment, and the compensation does not remove the effect on the duration of the coupled pulse. While the drop in efficiency depends on the central wavelength and duration of the pulse as well as the HCF core size and the window material, we find that the distance required to avoid significant Kerr-lensing effects can be found generally by keeping the maximum total nonlinear phase shift in the window (commonly referred to as the B-integral) below approximately 0.2.

\section{Numerical method}
Our simulations use our open-source propagation code Luna.jl~\cite{brahms_lunajl_2021} and are based on solving the unidirectional pulse propagation equation~\cite{kolesik_nonlinear_2004} in cylindrical symmetry:
\begin{equation}
  \partial_z E(\omega, k_\perp, z) = i\left(k_z(\omega, k_\perp) - \frac{\omega}{v_\mathrm{g}}\right)E(\omega, k_\perp, z) + i\frac{\mu_0 \omega^2}{2k_z}P^\mathrm{nl}(\omega, k_\perp, z)\,,
\end{equation}
where $z$ is propagation distance, $E(\omega, k_\perp, z)$ is the electric field in reciprocal space, $\omega$ is angular frequency, $k_\perp$ is the transverse component of the wave vector, $k_z(\omega, k_\perp)$ is the longitudinal component, $v_\mathrm{g}$ is the group velocity of the reference frame, $\mu_0$ is the vacuum permeability and $P^\mathrm{nl}(\omega, k_\perp, z)$ is the induced nonlinear polarisation. The real-space and reciprocal-space representations of the field $E$ are related by a Fourier transform in time and $0^\mathrm{th}$-order Hankel transform in transverse space:
\begin{align}
  E(\omega, k_\perp, z) &= \int_{-\infty}^\infty \dd t \int_{0}^{\infty} r \dd r E(t, r, z) J_0(k_\perp r) \mathrm{e}^{i\omega t}\\
  E(t, r, z) &= \frac{1}{2\pi}\int_{-\infty}^\infty \dd \omega \int_{0}^{\infty} k_\perp \dd k_\perp E(\omega, k_\perp, z) J_0(k_\perp r) \mathrm{e}^{-i\omega t}\,,
\end{align}
where $J_0$ is the $0^\mathrm{th}$-order Bessel function of the first kind. Similar equations apply to the polarisation $P^\mathrm{nl}$. The Hankel transform is implemented using the quasi-discrete Hankel transform algorithm~\cite{Yu1998} with  a pupil radius of \SI{15}{\mm} and 4096 samples. The longitudinal wave vector $k_z$ is given by
\begin{equation}
  k_z = \sqrt{\frac{\omega^2}{c^2}n^2(\omega) - k_\perp^2}\,,
\end{equation}
where $n(\omega)$ is the refractive index of the window material and $c$ is the speed of light. We only consider the effects of Kerr nonlinearity and dispersion in the window. To reduce the size of the computational grid, we use envelope (analytic) fields in the propagation and neglect third-harmonic generation. The nonlinear polarisation is hence given by
\begin{equation}
  P^\mathrm{nl}(t, r, z) = \frac{3}{4}\epsilon_0\chi^{(3)}\abs{E(t, r, z)}^2E(t, r, z)\,,
\end{equation}
where $\epsilon_0$ is the vacuum permittivity and $\chi^{(3)}$ is the third-order susceptibility of the window material.

The initial condition is a pulse with a Gaussian profile in time and space. The $1/\mathrm{e}^2$ radius is set to $w_0 = 0.64a$ with $a$ the HCF core radius, for which the coupling efficiency to the fundamental mode is optimal at \SI{98.07}{\percent}. We back-propagate this beam to the plane of the entrance window, forward-propagate through the window using the UPPE, and then forward-propagate to the nominal focal plane to find the pulse entering the HCF. Because we start with a uniform Gaussian beam in the focal plane, all spectral components of the pulse focus to the same spot size in the absence of nonlinearity. This is not the case for real laser beams, which are usually spatio-spectrally uniform when collimated rather than focused, but it allows us to directly compare the coupling efficiency to the optimal value without the need to re-calculate this for broadband beams. In addition, we add a spectral phase to the initial pulse before the window which perfectly compensates the dispersion of the window. This means that in the absence of nonlinearity, the pulse entering the HCF is transform-limited. To find the pulse which is coupled into the fundamental mode of the HCF, we calculate the overlap integral between the beam in the plane of the HCF and the mode:
\begin{equation}
  \label{eq:eta}
  E_m(t)= 2\pi\int_0^\infty\hat{\bm{e}}^*(r)\cdot \bm{E}(t, r, z_\textsc{HCF})\,r\dd r\,,
\end{equation}
where $\hat{\bm{e}}^*(r)$ is the normalised field distribution of the fundamental mode HE$_{11}$ as calculated using the capillary model~\cite{marcatili_hollow_1964}. While this assumption is only accurate for simple hollow capillary fibres, other hollow-core fibres can be closely approximated using this model~\cite{finger_accuracy_2014}. The ratio between the energy contained in $E_m(t)$ and the total pulse energy then gives the coupling efficiency.

\section{Coupling efficiency}
\begin{figure}
  \centering
  \includegraphics[width=3.4in]{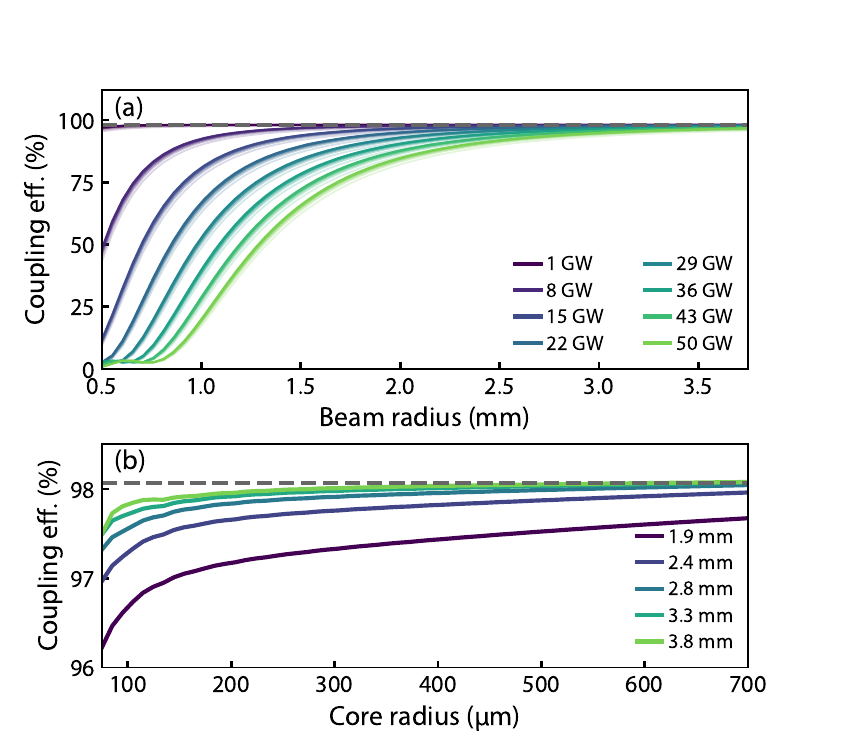}
  \caption{(a) Coupling efficiency to the fundamental mode of the HCF as a function of beam size in the window plane for \SI{10}{\fs} pulses at \SI{800}{\nm} and a range of peak powers (colours) and HCF core radii (line opacities). (b) Same as (a) but shown as a function of core radius for a fixed peak power of \SI{10}{\GW} and a range of beam radii (colours).}
  \label{fig:beamsize_radius}
\end{figure}

Figure \ref{fig:beamsize_radius}(a) shows the simulated coupling efficiency to an HCF as a function of beam size at the window for a \SI{10}{\fs} pulse centred at \SI{800}{\nm}, a peak power between \SI{1}{\GW} and \SI{50}{\GW}, and a \SI{1}{\mm} thick silica window. For each colour representing a peak power, we show several lines, which correspond to different HCF core radii from \SI{75}{\um} to \SI{265}{\um}, with darker lines corresponding to larger cores. Three main features can be observed in this data: the coupling efficiency increases for larger beam sizes, as expected from the lower intensity and hence weaker Kerr lens; the beam size required for high-efficiency coupling increases strongly for larger peak powers; and there is only a very weak dependence on core radius, as visible from the fact that the different lines at each colour overlap very closely. The latter observation is confirmed in Fig.~\ref{fig:beamsize_radius}(b), where we show the trend with core size over a larger range for a fixed peak power of \SI{10}{\GW} and a few different beam sizes. Despite a difference in core radius of around an order of magnitude, the coupling efficiency changes only very little. For the remainder of the simulations shown in this article, we fix the core radius at \SI{125}{\um}. Note that because the spot size of the focus is proportional to the core radius, the distance required to obtain a certain beam size at the window is different for each core radius. We calculate the required distance from the parameters of the initial Gaussian beam. Defining the beam size rather than the distance itself in our simulations ensures that the far-field beam always fits well within our radial grid, avoiding edge artefacts.

\begin{figure}
  \centering
  \includegraphics[width=5.25in]{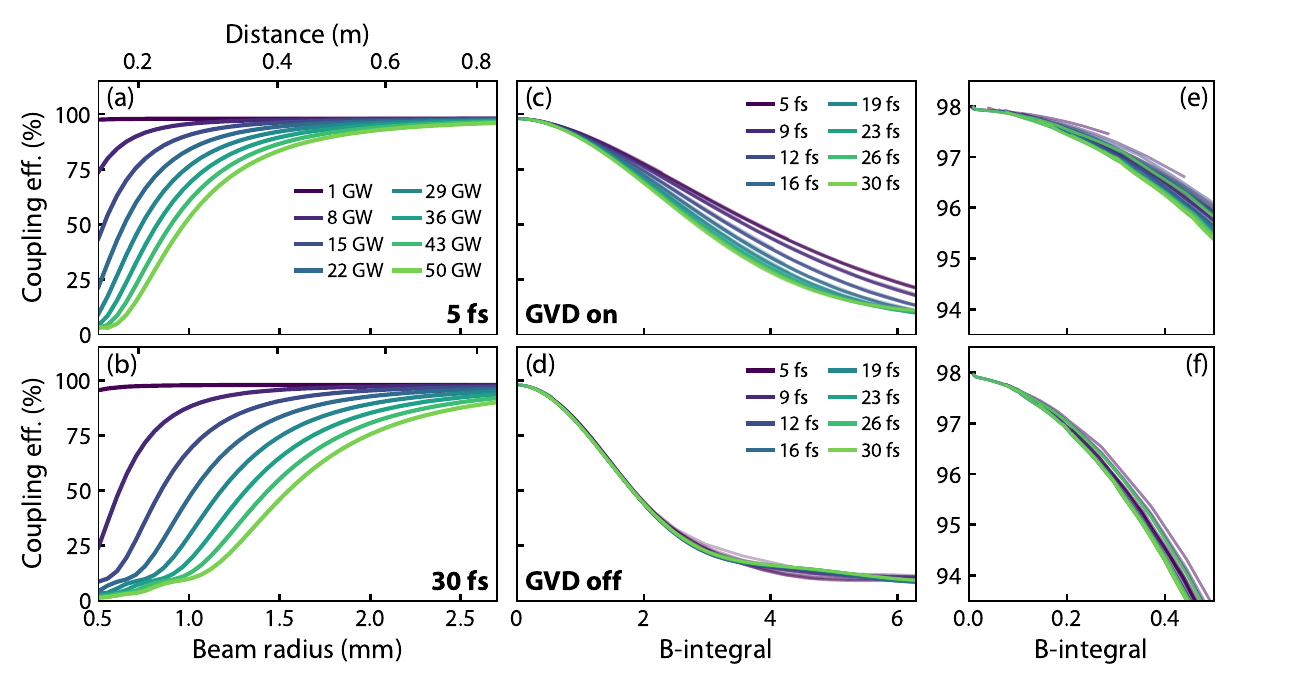}
  \caption{(a) Coupling efficiency to the fundamental mode of an HCF with \SI{125}{\um} core radius as a function of beam size in the window plane for nominally \SI{5}{\fs} pulses at \SI{800}{\nm} and the same peak powers as shown in Fig.~\ref{fig:beamsize_radius}(a). The top axis shows the distance between the HCF entrance face and the window. (b) same as (a) but for a nominal pulse duration of \SI{30}{\fs}. (c) Coupling efficiency as a function for a range of pulse durations for otherwise identical conditions. For each pulse duration, several lines corresponding to different peak powers are overlaid. (d) zoomed-in view of the same data shown in (c). (e) and (f), same as (c) and (d) but when neglecting the GVD of the window in the simulations.}
  \label{fig:scan_duration}
\end{figure}

The results shown in Fig.~\ref{fig:beamsize_radius} are for pulses with a nominal duration of \SI{10}{\fs}. In Figs.~\ref{fig:scan_duration}(a) and (b) we compare the results for two different pulse durations, \SI{5}{\fs} and \SI{30}{\fs}. It can be clearly seen that for shorter pulses, the beam at the entrance window can be smaller while still maintaining good coupling efficiency. This is due to the fact that the initial pulse impinging on the window is chirped to pre-compensate for the window dispersion, which has a stronger effect on shorter pulses. In the examples shown here, the \SI{5}{\fs} pulse is pre-chirped to \SI{19}{\fs}, a nearly four-fold increase in duration, while the duration of the \SI{30}{\fs} pulse remains approximately the same. For the same \emph{nominal} peak power, the \emph{actual} peak power is hence lower for shorter pulses. To verify that this is the dominant effect when changing the pulse duration, we compare the results across a range of pulse durations and peak powers. To show the results on a unified scale, we calculate the total nonlinear phase shift at the centre of the beam under the assumption that the beam does not change shape significantly during propagation through the window:
\begin{equation}
  B = n_2 k_0 \int_0^L I_0(z) \dd z = \frac{2n_2 k_0}{\pi w^2} \int_0^L P_0(z) \dd z\,,
  \label{eq:Bint}
\end{equation}
where $n_2$ and $L$ are the nonlinear refractive index and the thickness of the window material, respectively, $w$ is the beam radius at the window, $k_0$ is given by $k_0=2\pi/\lambda$ with $\lambda$ the pulse wavelength, and $I_\mathrm{0}$ and $P_\mathrm{0}$ are the peak intensity and peak power of the pulse, respectively. We calculate $P_0(z)$ by numerically propagating the initial pre-chirped pulse through the window considering only the window dispersion. For simplicity, we also assume that the on-axis pulse has the nominal pulse shape we expect to couple into the HCF, that is, we neglect the spatio-spectral coupling caused by the difference in divergence across the pulse spectrum. The results are shown in Fig.~\ref{fig:scan_duration}(c). For each pulse duration, we overlay the results for the same range of nominal peak powers. From the fact that these lines overlap exactly, it is clear that the coupling efficiency remains the same when changing the peak power but simultaneously adjusting the beam size to keep the intensity fixed. The difference in coupling efficiency for different pulse durations is much less pronounced when presented in this way, suggesting that for the same total nonlinearity (B-integral), the effect on the focusing conditions is similar. For comparison, Fig.~\ref{fig:scan_duration}(d) shows the same but when neglecting the window dispersion entirely, both in the propagation simulations and in calculating the B-integral. In this case, there is virtually no difference between pulse durations. By zooming in to the region closest to zero nonlinearity, as shown in Figs.~\ref{fig:scan_duration}(e) and (f), we see that the results for different pulse durations become virtually indistinguishable in the region below a B-integral of around 0.4. Conservatively, we can conclude that keeping the B-integral in the window below 0.2 reduces the effect of the Kerr lens to be negligible regardless of core size, pulse duration, or peak power. Note that some of the lines shown in Fig.~\ref{fig:scan_duration} do not extend to the region of lowest B-integral. This is because the maximum beam radius in our simulations is set by the size of our computational grid; for the highest peak powers, the maximum beam size is insufficient to reduce the nonlinearity to this extent. 

\begin{figure}
  \centering
  \includegraphics[width=5.25in]{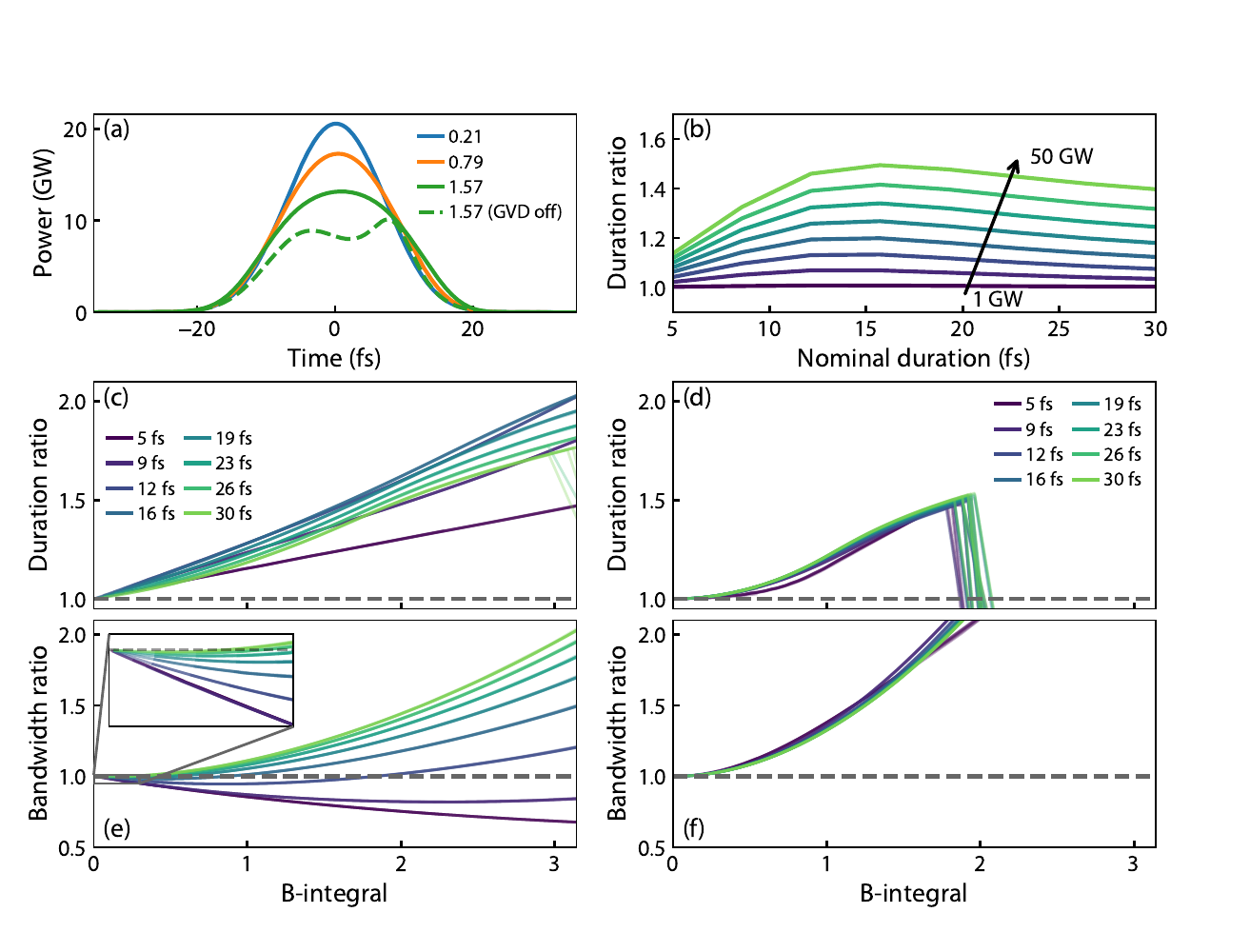}
  \caption{(a) Examples of coupled pulses for an initial pulse duration of \SI{16}{\fs} and three different values of the B-integral. (b) Ratio between coupled and initial FWHM pulse duration for a beam size at the window of \SI{2}{\mm} as a function of nominal pulse duration for a range of nominal peak powers. (c) Ratio between coupled and initial FWHM pulse duration as a function of B-integral for a range of pulse durations. (d) Same as (c) but with the window dispersion neglected. (r) Ratio between coupled and initial RMS frequency bandwidth. (f) Same as (e) but with the window dispersion neglected.}
  \label{fig:pulse_stretching}
\end{figure}

\section{Pulse stretching}
In addition to reducing the amount of energy coupled into the HCF, the nonlinearity of the window also affects the shape of the pulse. The peak of the pulse creates a stronger nonlinear lens than the wings and the wavefront at the pulse peak is thus more strongly distorted. The effect is equivalent to a time-dependent attenuation which removes energy from the peak of the pulse and leads to an effective pulse stretching. In addition, because pre-compensating the dispersion of the window makes the pulse negatively chirped, self-phase modulation leads to spectral compression~\cite{oberthaler_special_1993}. Rather than emerging from the window at its transform-limited shape in time, the pulse is therefore already somewhat stretched. Figure~\ref{fig:pulse_stretching}(a) shows an example of these effects for a nominal pulse duration of \SI{16}{\fs}, a peak power of \SI{22}{\GW} and three different window distances, resulting in different values of the B-integral. As the nonlinearity increases, the peak of the pulse is attenuated, but the weaker wings are less strongly affected, and hence the pulse duration increases. The green dashed line shows the result for the example with the highest B-integral but with the window dispersion neglected, isolating the spatio-temporal effects. Here the peak of the pulse is even more strongly attenuated, because the pulse is fully temporally compressed throughout the propagation through the window.

Figure~\ref{fig:pulse_stretching}(c) shows the ratio between the full-width-at-half-maximum (FWHM) durations of the coupled pulse $E_m(t)$ and the input pulse for the same range of pulse durations shown in Fig.~\ref{fig:scan_duration}. For values of the B-integral of up to around $\pi$, the coupled pulse is longer than the input for all conditions shown, with an approximately linear trend. Above this value, the pulse is strongly distorted and shorter features appear, leading to a reduction in FWHM duration; however, this regime is of limited practical interest, because the coupling efficiency is low. Note that in Fig.~\ref{fig:pulse_stretching}(c), results for different peak powers are once again overlaid in the plot, but are very close together and indistinguishable on this scale. Importantly, the pulse-stretching effect is very small for a B-integral below 0.2, with the coupled pulse only a few percent longer than its nominal duration---in good agreement with the results for the coupling efficiency. In contrast to the coupling efficiency, the pulse stretching depends strongly on the nominal pulse duration. The shortest pulse is least affected, followed by the longest. Pulses with durations of \SI{12}{\fs} and \SI{16}{\fs} experience the largest increase in their duration, as shown in Fig.~\ref{fig:pulse_stretching}(b). This trend is likely due to the interplay between the spatio-temporal distortion and the spectral-domain effects of the nonlinearity; a conclusion which is supported by the fact that the trend disappears when the window dispersion is neglected, as shown in Fig.~\ref{fig:pulse_stretching}(d).

Fig.~\ref{fig:pulse_stretching}(e) shows the ratio between coupled and nominal root-mean-square frequency widths from the same simulations. For low values of the B-integral (see inset), the coupled bandwidth is universally lower, because weak self-phase modulation leads to spectral compression only. For stronger nonlinearity, the effect depends on the pulse duration: long pulses are only weakly negatively chirped before the window and hence quickly reach the point of approximately zero chirp, after which they spectrally broaden, whereas short pulses spectrally compress even for large B-integral values. This trend also disappears when the window dispersion is neglected [see Fig.~\ref{fig:pulse_stretching}(f)]. Here, pulses of all durations experience only spectral broadening, because the initial pulses are transform-limited.

\section{Influence of wavelength and window material}
\begin{figure}
  \centering
  \includegraphics[width=5.25in]{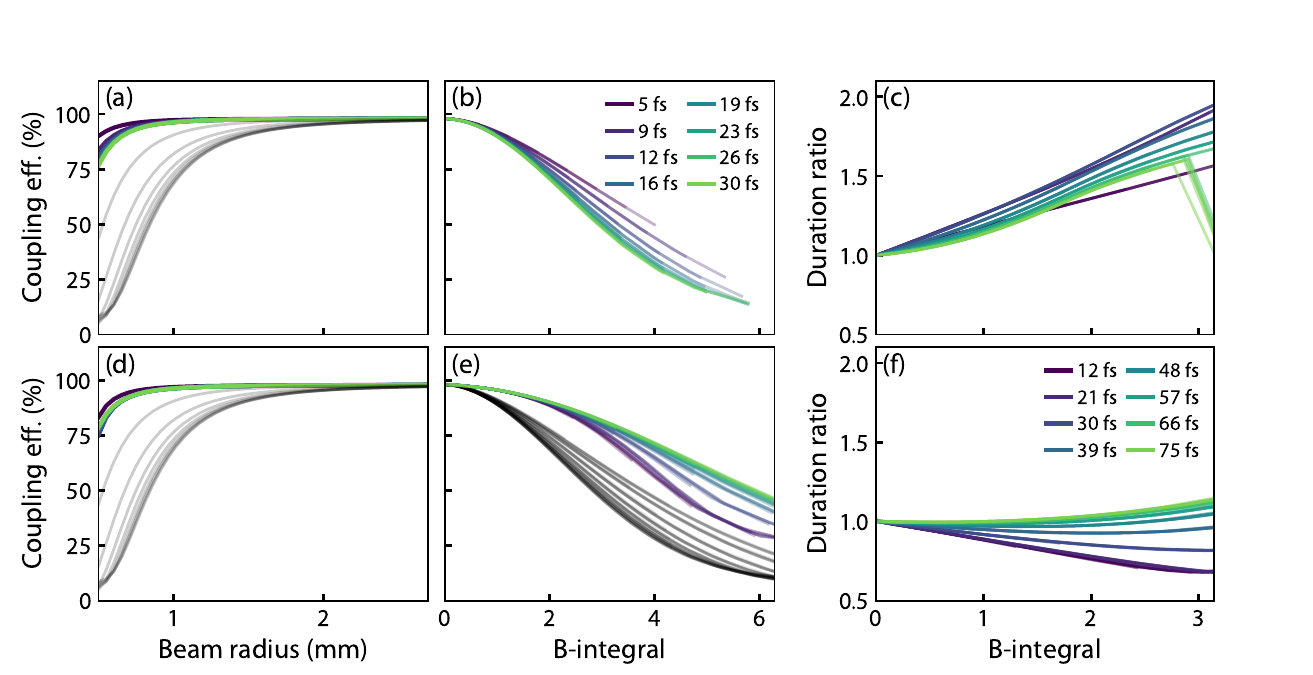}
  \caption{(a) Coupling efficiency as a function of beam radius at the window for \SI{15}{\giga\watt} nominal peak power at \SI{800}{\nm} for a \SI{1}{\mm} thick MgF$_2$ window. Each colour corresponds to a pulse duration. The grey lines show the equivalent results for a \SI{1}{\mm} thick silica window. (b) Coupling efficiency as a function of B-integral. Results for different peak powers are overlaid with semi-transparent lines. (c) Ratio of coupled to nominal pulse duration for the same conditions as in (a) and (b). (d) to (f) Same as (a) to (c) but for a \SI{1}{\mm} thick silica window and pulses centred at \SI{2000}{\nm}. The grey lines in (e) show the results for pulses at \SI{800}{\nm} for comparison. The legend in (b) applies to panels (a) to (c); the legend in (f) applies to panels (d) to (f).}
  \label{fig:MgF2_and_IR}
\end{figure}
The results so far were obtained with pulses centred at \SI{800}{\nm} and silica windows. Among commonly used window materials, magnesium fluoride (MgF$_2$) exhibits one of the lowest nonlinear refractive indices: \SI{5.8e-17}{\square\cm\per\watt} as compared to \SI{2.6e-16}{\square\cm\per\watt} for silica glass~\cite{desalvo_infrared_1996}. For the same thickness, a MgF$_2$ window should therefore allow for much higher intensity before significantly disturbing the coupling conditions. Fig.~\ref{fig:MgF2_and_IR}(a) confirms that this is indeed the case. The coloured lines show the simulated coupling efficiency for a \SI{1}{\mm} thick MgF$_2$ window and \SI{15}{\giga\watt} nominal peak power, while the grey lines show the same for a silica window; both sets of data cover the same range of pulse durations. For any given beam radius at the window, the MgF$_2$ window has a much smaller effect. This is despite the fact that the lower group-velocity dispersion of MgF$_2$ (\SI{20.5}{\fs\squared\per\mm} for MgF$_2$~\cite{li_refractive_1980} compared to \SI{38.6}{\fs\squared\per\mm} for silica~\cite{palik_handbook_1985} at \SI{800}{\nm}) leads to higher actual peak power inside the window for the same nominal peak power. When plotted against the B-integral [Fig.~\ref{fig:MgF2_and_IR}(b)], the trends from the simulations using silica are generally reproduced; the results using silica are not shown for comparison as they overlap closely with those for MgF$_2$. However, the data does not extend to the same range of total nonlinearity, because the range of beam size we consider is the same and the nonlinearity for a given beam size is lower for MgF$_2$. Fig.~\ref{fig:MgF2_and_IR}(c) shows the pulse stretching for the same range of parameters. Here, the main difference to the results in silica is the closer similarity between different pulse durations. This is a result of the lower group-velocity dispersion in the window; the results using MgF$_2$ lie in between those for silica [Fig.~\ref{fig:pulse_stretching}(c)] and those neglecting the window dispersion [Fig.~\ref{fig:pulse_stretching}(d)].

Moving to longer wavelengths in the infrared also affects the self-focusing behaviour. Figures \ref{fig:MgF2_and_IR}(d), (e) and (f) show results obtained with pulses at \SI{2000}{\nm} central wavelength and a silica window. The range of nominal pulse duration is scaled to keep the number of cycles under the envelope the same as at \SI{800}{\nm}. As in Fig.~\ref{fig:MgF2_and_IR}(a), the grey lines repeat the results at \SI{800}{\nm} for comparison. The effect of self-focusing is significantly weaker. One aspect contributing to the difference is that the magnitude of the dispersion of silica is larger at \SI{2000}{\nm} (\SI{-87.6}{\fs\squared\per\mm}) than at \SI{800}{\nm}, so the actual peak power at the entrance face of the window is more strongly reduced (this effect is counteracted by the longer pulse duration for longer wavelengths, which reduces the effect of dispersion on the pulse duration). However, the difference remains when taking the initial chirp into account by plotting the results against the B-integral, as shown in Fig.~\ref{fig:MgF2_and_IR}(e), and also when ignoring the window dispersion entirely (not shown). The dispersion alone does thus not explain why longer wavelengths are less strongly affected. The key difference lies in the wavefront curvature of the beam entering the window. Absent any nonlinearity, the beam waist is located at the HCF entrance, and hence the radius of curvature of the wavefront at the window is approximately equal to the distance between the HCF and the window, assuming the window is much further away from the HCF than the Rayleigh length. For a fixed HCF core size, the beam at a longer wavelength reaches a given beam size closer to the HCF entrance face because diffraction is stronger, and consequently its wavefront is more strongly curved. Therefore, the additional curvature caused by the Kerr lens---which, for a fixed window and neglecting dispersion, depends only on the beam size and peak power---has a smaller effect on the subsequent focusing of the beam. As a simple model, we can consider a collimated beam impinging on an effective focusing element which consists of two optics: a perfect lens with focal length equal to the window distance and the nonlinear lens. The effective focal length is then given by
\begin{equation}
  \frac{1}{f_\mathrm{eff}} = \frac{1}{d_\mathrm{win}} + \frac{1}{f_\mathrm{Kerr}}\,,
\end{equation}
where $d_\mathrm{win}$ is the distance from the window to the HCF and $f_\mathrm{Kerr}$ is the focal length of the Kerr lens. The latter can be defined in several ways~\cite{leghmizi_different_2017}, the simplest of which is the parabolic approximation to a Gaussian beam profile, which results in
\begin{equation}
  f_\mathrm{Kerr} = \frac{\pi w^4}{8n_2LP_0}\,,
\end{equation}
where $w$ is the beam radius in the nonlinear medium. When $w$ and $P_0$ are fixed (which corresponds to a specific point on one of the lines in Fig.~\ref{fig:MgF2_and_IR}(d)), the Kerr focal length does not depend on the wavelength, but $d_\mathrm{win}$ does, and hence so does the effective focal length and the shift in the coupling conditions. The effect is significant but cannot be captured by a simple scaling rule, such as a scaling of the B-integral by the wavelength. The limiting value of the B-integral of 0.2 is thus still useful but more conservative for longer wavelengths. 

Because silica, like nearly all optical materials, exhibits anomalous dispersion at \SI{2000}{\nm} wavelength, the pulse is positively chirped to pre-compensate the window dispersion. This has a dramatic effect on the shape of the pulse coupled into the waveguide. As shown in Fig.~\ref{fig:MgF2_and_IR}(f), the coupled pulse is almost universally shorter than its nominal duration, with only very high B-integral values leading to pulse stretching. This effective compression also disappears when the window dispersion is switched off. We attribute this to the combination of two effects: firstly, as demonstrated by the higher coupling efficiency in Fig.~\ref{fig:MgF2_and_IR}(e), the long-wavelength pulse is less strongly affected by the Kerr lens, and hence it is likely that the amount of spatio-temporal reshaping of the pulse is also reduced. Secondly, because it is initially positively chirped, the pulse experiences spectral broadening rather than spectral compression, and is simultaneously temporally compressed by the anomalous dispersion of window. Because this can occur without a significant reduction in the coupling efficiency, the peak power of the coupled pulse can be higher than its nominal value.

\section{Compensation by moving the focusing optic}
\begin{figure}
  \centering
  \includegraphics[width=3.4in]{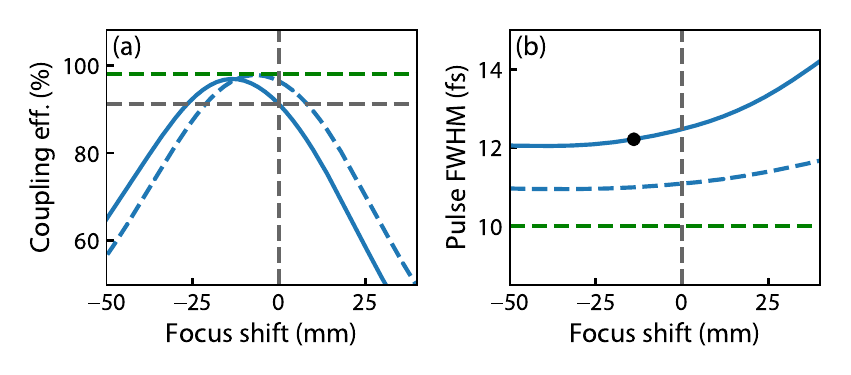}
  \caption{(a) Coupling efficiency for \SI{50}{\giga\watt}, \SI{10}{\fs} pulses at \SI{800}{\nm} when compensating for the Kerr-lens induced reduction in coupling efficiency by shifting the nominal focus. The window is \SI{1}{\mm} thick silica and the beam radius is \SI{2.5}{\mm}. The blue dashed line shows the same but for a lower peak power of \SI{22}{\giga\watt}. The green dashed line shows the optimal coupling efficiency of 98.07\%. The horizontal grey dashed line shows the coupling efficiency without any compensation. (b) Coupled pulse duration for the same conditions as in (a). The green dashed line shows the nominal pulse duration. The black dot shows the pulse duration at the position of maximum coupling efficiency in (a).}
  \label{fig:compensation}
\end{figure}

Because the main spatio-temporal effect of the nonlinearity in the window is an effective positive lens, the reduction in coupling efficiency to the HCF can be counteracted at least in part by shifting the nominal focus of the coupling optics to lie beyond the HCF entrance facet. The Kerr lens then causes additional wavefront curvature and shifts the focus to the correct plane. This has previously been applied experimentally to compensate the nonlinear effects of not only the entrance window but also additional transmissive optics in the beam path~\cite{ran_coupling_2020} and is a common practical step in ultrafast laser laboratories using HCF. Fig.~\ref{fig:compensation}(a) shows how the coupling efficiency changes as the nominal focus is shifted to compensate for the nonlinear lens. As expected, reducing the distance between the final focus optic and the HCF entrance facet increases the coupling efficiency. In this example, shifting by \SI{14}{\mm} leads to an improvement from 91.1\% to 96.8\% efficiency. The optimal efficiency in the absence of nonlinearity cannot be fully recovered. More importantly, the temporal reshaping of the pulse is barely affected by the compensation: as shown in Fig.~\ref{fig:compensation}(b), the pulse duration, which is stretched to around \SI{12}{\fs} in this case, improves only marginally. This effect is less pronounced for longer pulses but present in all examples we have considered.

Apart from the weak improvement in pulse duration, the main limitation on this compensation strategy is the dependence on input peak power. As shown by the dashed blue lines in Fig.~\ref{fig:compensation}, a different focus shift is required to optimise the coupling efficiency for a different peak power---in this case, a peak power of \SI{22}{\GW} requires a shift of \SI{6}{\mm} instead of \SI{14}{\mm}. As a result, for a fixed focusing geometry, changing the input pulse energy can dramatically change the coupling efficiency. In addition, the duration of the coupled pulse changes [see dashed blue line in Fig.~\ref{fig:compensation}(b)]. For systems in which the pulse energy is kept constant under normal operating conditions (for instance, HCF used for pulse compression via spectral broadening and post-compression~\cite{nisoli_compression_1997}), this limitation may be acceptable and higher intensity on the window can be tolerated, making such systems more compact. However, many more advanced light sources based on gas-filled HCF require precise adjustment of the pulse energy to achieve the desired result~\cite{travers_high-energy_2019,kottig_efficient_2020,mak_two_2013,lekosiotis_generation_2020}. In addition, some techniques in this category require short driving pulses~\cite{travers_high-energy_2019,brahms_high-energy_2019}. For these systems to operate reliably and without major distortions to the input pulse shape, the nonlinearity must be kept low and compensating by shifting the focus is less useful.

\section{General design rules}
\begin{figure}
  \centering
  \includegraphics[width=4.5in]{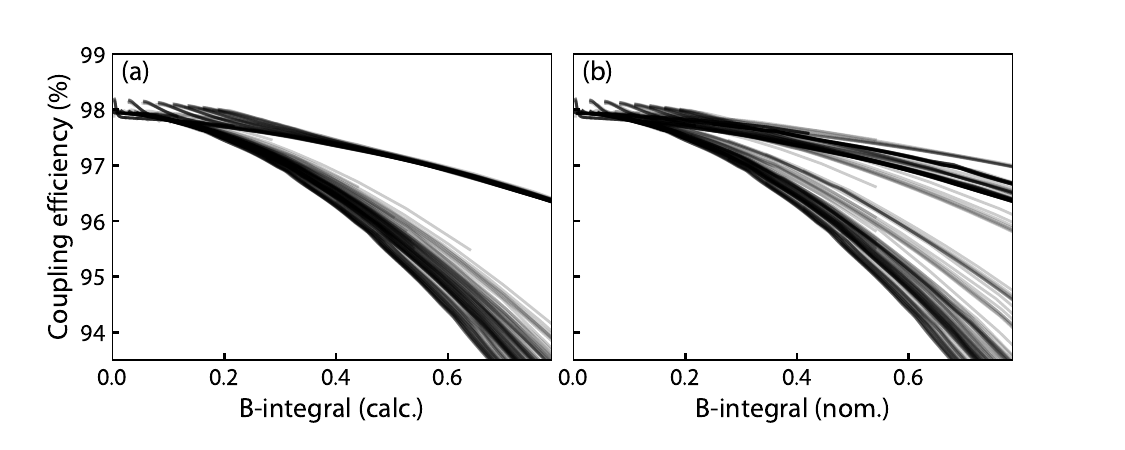}
  \caption{(a) Overlay of all results for the coupling efficiency obtained with a variety of pulse durations, peak powers, window materials, and central wavelengths, plotted against the B-integral as calculated via eq.~\ref{eq:Bint}. (b) Same as (a) but plotted against the B-integral as calculated assuming the nominal pulse duration and constant peak power.}
  \label{fig:coupling_all}
\end{figure}

In the results shown so far, keeping the nonlinearity below a maximum B-integral value of around $0.2$ is sufficient to keep the effects of the Kerr lens to a minimum. However, calculating the B-integral using eq.~\ref{eq:Bint} is relatively cumbersome as it requires numerical integration of the linear pulse propagation through the material. In Fig.~\ref{fig:coupling_all} we compare the results when plotting against the B-integral calculated in two ways: as using eq.~\ref{eq:Bint}; or by simply ignoring the dispersion altogether and assuming the nominal peak power and duration are maintained throughout the window. For an overview, we overlay all of our results in semi-transparent lines. In addition to the results shown so far, we include another set of simulations for longer pulses (\SI{30}{\fs} to \SI{300}{\fs}) centred at \SI{1030}{\nm} to correspond to the parameters of ytterbium-based laser systems; there are no significant new features in this data which have not been discussed already. While there are significant differences in the individual simulation results, the overall conclusion on the maximum B-integral remains the same, regardless of how we calculate the B-integral. Since the limit on the B-integral does not strongly depend on the pulse parameters, the required distance between the HCF entrance face and the window can be calculated by simply finding the distance at which the beam has expanded sufficiently to drop the B-integral below the limit. Assuming the window is always placed far from the focus, and hence well outside the Rayleigh length, the distance can be calculated as
\begin{equation}
  d_\mathrm{min} = 2\pi \left[ \frac{n_2 L P_0 x_a^2 a^2}{\lambda_0^3 B_\mathrm{max}} \right]^{\frac{1}{2}} \approx 4a \left[ \frac{n_2 L P_0}{\lambda^3 B_\mathrm{max}} \right]^{\frac{1}{2}}\,,
\end{equation}
where $P_0$ is the nominal peak power of the pulse, $B_\mathrm{max}$ is the maximum B-integral, and $x_a = 0.64$ is the ratio between the beam radius and the core radius.

\section{Conclusions}
In summary, we have numerically investigated the effect of nonlinear lensing in gas-cell entrance windows on the coupling of ultrafast laser pulses to hollow-core fibres. We have found that the nonlinear influence on the focusing degrades the coupling efficiency and, in most cases, leads to an effective stretching of the coupled pulse due to both nonlinear spectral compression in the window and spatio-temporal reshaping. For pulses which are nominally chirp-free on the far side of the window, the interplay between the window dispersion and nonlinear spatio-temporal effects causes significant differences in the behaviour for different window materials, pulse durations, and pulse wavelengths. In addition, the stronger effect of diffraction on longer-wavelength beams means they are less susceptible to nonlinear lensing higher intensity can be used. The core size of the hollow-core fibre plays only a minor role. The largest differences are found in the temporal reshaping of the pulse, with some conditions leading to pulse compression and others to stretching. While the coupling efficiency can be restored to a large extent by shifting the nominal focal plane to compensate for the Kerr lens, the effect on the pulse duration cannot be removed in this way. Our most important finding, however, is that a relatively simple limit on the B-integral, as calculated using the \emph{nominal} pulse parameters, is applicable to all parameter combinations we have tested. By choosing the beam size according to this limit, nonlinear lensing effects can be kept to a minimum, ensuring the efficient and distortion-free delivery of ultrafast laser pulses to the waveguide.

\begin{backmatter}
\bmsection{Funding} This work was funded by the Royal Academy of Engineering under the Research Fellowship Programme, grant agreement RF/202122/21/133.

\bmsection{Acknowledgments}
Thanks to J.C. Travers for useful discussions.

\bmsection{Disclosures} The authors declare no conflicts of interest.

\bmsection{Data availability}
The data that support the findings of this study are available from the corresponding author upon reasonable request.

\end{backmatter}

\bibliography{bibliography}

\end{document}